\documentclass{aa}  

\usepackage{graphicx}
\usepackage{txfonts}
\usepackage{hyperref}
\hypersetup{
  colorlinks   = true, 
  urlcolor     = blue, 
  linkcolor    = blue, 
  citecolor   = blue 
}
\usepackage{orcidlink}
\usepackage{xcolor}
\newcommand{\review}[1]{\textcolor{black}{#1}}
\newcommand{\highlight}[1]{\textcolor{black}{#1}}

\newcommand\romanst{\textit{Roman}}

\newcommand{\ari}{Zentrum f{\"u}r Astronomie der Universit{\"a}t Heidelberg, Astronomisches Rechen-Institut, M{\"o}nchhofstr. 12-14, 69120 Heidelberg, Germany}
\newcommand{\llnl}{Space Science Institute, Lawrence Livermore National Laboratory, 7000 East Ave., Livermore, CA 94550, USA}
\newcommand{\git}{Georgia Institute of Technology, 837 State St., Atlanta, GA 30309, USA}

\begin{document}

   \title{Astrometric microlensing probes of the isolated neutron star population with \textit{Roman}}

   \subtitle{}

   \author{Z. Kaczmarek\orcidlink{0009-0007-4089-5012}\inst{1,3}\fnmsep\thanks{\email{zofia.kaczmarek@uni-heidelberg.de}}
          \and
          A. Halasi-Kun\orcidlink{0009-0007-4473-8321} \inst{2,3}
          \and
          P. McGill\orcidlink{0000-0002-1052-6749} \inst{3}
          \and
          S.E. Perkins\orcidlink{0000-0002-5910-3114} \inst{3}
          \and
          W.A. Dawson\orcidlink{0000-0003-0248-6123} \inst{3}
          }

   \institute{
        \ari
         \and
         \git
         \and
         \llnl
             }

   \date{Received XXX; accepted YYY}

   \abstract
   {Notoriously hard to detect and study, isolated neutron stars (NS) could provide valuable answers to fundamental questions about stellar evolution and explosion physics. With the upcoming \textit{Roman Space Telescope}, scheduled for launch in 2026, a new and powerful channel for their detection --  astrometric microlensing -- will become available.}
   {We set out to create a realistic sample of simulated gravitational microlensing events as observed by \textit{Roman} with the Galactic Bulge Time Domain Survey. We focus in particular on the population of NS lenses, which has until now been largely understudied.}
   {We use state-of-the-art Galactic models tailored for application to microlensing by compact objects. In addition to populations of stars, white dwarfs and black holes, we simulate four different NS populations with Maxwellian natal kick distributions: $\bar{v} = (150, 250, 350, 450)$ km/s. For each simulation, we apply projected \textit{Roman} precision, cadence, and detectability criteria.}
   {We find the parameter space $\log_{10} t_{\rm E}$ -- $\log_{10} \theta_{\rm E}$, which will be accessible to \textit{Roman} observations, to be maximally efficient for classification of stellar remnants. We find a feature in this space that is characteristic to NS; using this feature, optimal samples of NS candidates can be constructed from \textit{Roman}-like datasets.
   We describe the dependence of observable parameter distributions on the assumed mean kick velocities. As the effects of natal kicks are very complex and mutually counteracting, we suggest more detailed studies focused on the dynamics of NS are needed in anticipation of \textit{Roman} and future surveys. We estimate \textit{Roman} will observe approximately 11000 microlensing events -- including $\sim$100 with NS lenses -- whose both photometric and astrometric signal are detectable; the event yield decreases by 38\% if gap-filling low-cadence observations are not included. We make all simulated microlensing event datasets publicly available in preparation for \romanst~data.}
   {}

   \maketitle

\section{Introduction}
\label{sec:intro}

Neutron stars (NS) are among the most extreme objects in the present-day Universe \citep{Annala2023}, with core densities greatly exceeding the nuclear saturation limit \citep{Haensel2007, Koehn2025}. Since such environments cannot be reproduced on Earth, NS are natural laboratories for studying matter in extreme pressure and density conditions \citep{Lattimer2004, Haensel2007}. The equation of state (EoS) describing NS is not yet accurately known \citep{Malik2024, Chatziioannou2024, Ji2025}; its applications reach far beyond astrophysics and into fundamental physics, in particular nuclear physics and quantum chromodynamics \citep[e.g.,][]{Baym2018, Lattimer2021, Kumar2024}. As allowed by some EoS, NS are speculated to contain a phase transition to exotic phases of matter \citep{Lattimer2004, Annala2020, Annala2023}, and astrophysical observations could in the near future be decisive in confirming or rejecting this hypothesis \citep{Somasundaram2023}. Since the EoS defines physical properties of NS -- including the mass-radius relation, minimum and maximum masses, and the overall mass distribution -- it is strongly connected to astrophysical observations and can be constrained by them \citep{Lattimer2012, Fraga2016, Baym2018, Ji2025}.

Among the great unknowns about NS is also the mechanism through which they are born. Although it has long been known that NS experience large natal kicks \citep{Gunn1970, LyneLorimer1994, Tauris1999, Hobbs2005}, their velocity distribution is still highly uncertain, and mean kick velocities varying between 100 and 500 kms$^{-1}$ are assumed \citep[e.g.,][]{Kalogera1998, Hobbs2005, BrayElridge2016, Igoshev2020, Igoshev2021, Fortin2022, Kapil2023, ODoherty2023, Disberg2025}. Several physical mechanisms for the origin of kicks have been proposed, resulting in distinctly different kick distributions \citep[e.g.,][]{Lai2004, Janka2017, Janka2024}. Verifying theoretical predictions for natal kicks is important for our understanding and modelling of supernovae \citep{Podsiadlowski2005, Mandel2020, Janka2024}. Population synthesis codes are reliant on natal kick prescriptions, which impact their predictions of gravitational wave observables \citep{Giacobbo2020} and interpretations of the nature of observed sources \citep{Zevin2020}.

The mass distribution of NS is also unknown, with the putative "mass gap" observed between high-mass NS and low-mass black holes \citep[][]{Bailyn1998, Ozel2010, Farr2011, Shao2022, ElBadry2024} presenting an open problem. It is unclear whether this mass gap is caused by an astrophysical mechanism \citep[e.g.,][]{Belczynski2012} or observational biases \citep[e.g.,][]{Wyrz2020}. While numerous isolated NS have been found through their pulsar emission \citep{pulsar_discovery, Bell2017}, NS masses can only be measured if they are in binaries, whether via observations of gravitational wave mergers \citep{NSmerger1, NSmerger2, BHNSmergers}, X-ray binaries \citep[e.g.,][]{Bhattacharyya2010, Rawls2011, Kim2021}, pulsar timing \citep[e.g.,][]{Taylor1979, Reardon2016} or astrometric wobble of the companion \citep{ElBadry2024}\footnote{\citet{Ono2015} have developed a method of mass measurement for rapidly rotating isolated NS via gravitational wave phase shift. However, it has not been applied yet and is beyond the sensitivity of all currently existing gravitational wave observatories.}. Such NS only represent rigidly defined evolutionary tracks and may not constitute a fully representative sample (e.g., \citealt{ElBadry2024} expect their NS sample to have originated in primordial binaries which survived the supernova explosions, and to have undergone very low kicks). NS masses and kicks are correlated and both depend on the progenitor mass \citep{Stone1982, Mandel2020}; kicks are also dependent on binarity \citep{Podsiadlowski2005}. Furthermore, being part of a binary system can also change NS parameters through mass transfer: the fastest-rotating and most massive known pulsar is a striking example of this effect \citep{Fonseca2021, Romani2022}. In summary, the sample of NS with measured masses is subject to complex selection effects.

This complicated interplay between evolutionary tracks and physical parameters of compact objects could be made clearer with the help of gravitational microlensing. This powerful technique of studying our Galaxy can be used to discover inherently dark objects, including compact objects \citep{Paczynski1996}. As microlensing requires a very close chance alignment of a light-deflecting mass and a luminous source, it is very rare; continuous monitoring of millions of stars -- typically in the densest regions of the sky -- is needed to secure detections \citep[e.g.,][]{Paczynski1991}. 
Hence, the potential of this technique has been realised on a large scale with specialised variability surveys \citep[e.g.,][]{OGLEIV, Sumi2013, KMTNet}, which nowadays publish databases of $\sim$10,000 microlensing events \citep[\review{at a rate of thousands of events per year}, e.g.,][]{Mroz2020, Husseiniova2021, Shin2024}, including events caused by stellar remnants. Microlensing has recently yielded the first confirmed discovery of an isolated stellar-origin black hole, OB110462 \review{\citep{Sahu2022, Lam2022, Mroz2022}}, setting the track for future population studies \citep{Lam2023, Perkins2024}. A recent analysis of a white dwarf by \citet{McGill2023} shows how a microlensing mass measurement can verify theoretical predictions for compact object structure and composition.

As of today, neutron stars still evade high-confidence microlensing detections. Despite extensive searches, no upcoming alignments of known pulsars with background stars close enough to cause a measurable microlensing signal have been found, as the pulsar sample is simply too small \citep{SchwarzSeidel2002, Ofek2018, Harding2018, Lu2024}. On the other hand, while numerous previously unknown dark lenses are found in variability surveys, insufficient constraints on mass prevent their confident classification as neutron stars. Within bounds allowed by the available data, the NS candidate events could also be explained by white dwarf, black hole or stellar lenses \citep[e.g.,][]{Wyrz2016, Wyrz2020, Lam2022_sup, Kaczmarek2022}. \review{In recent work, we have demonstrated that it is also virtually impossible to construct a useful sample of prospective NS lenses for astrometric follow-up from photometric information alone, as they occupy a similar region of the photometric observable parameter space as the far more common luminous stellar lenses \citep{Kaczmarek2025}.}

In this work, we investigate how simultaneous photometric and astrometric observations can overcome this limitation and provide valuable information on the NS population. Specifically, we focus on the upcoming \textit{Roman Space Telescope} \citep{Spergel2015}, the first space mission to include microlensing as a key science case. \romanst~will conduct the Galactic Bulge Time Domain Survey \citep[GBTDS;][]{Penny2019}, observing a $\approx 1.7 \deg^2$ region with a 12-minute cadence over six 70.5-day seasons\footnote{Roman Observations Time Allocation Committee Final Report and Recommendations; \url{https://roman.gsfc.nasa.gov/science/ccs/ROTAC-Report-20250424-v1.pdf}. Gaps between seasons will be filled with low-cadence observations.}. Those near-continuous observations, each yielding high-precision photometric (10 mmag) and astrometric (1 mas) measurements, will provide data of unprecedented value for studying microlensing events. Its applications range from collecting large \review{homogeneous} samples of planetary lenses for a statistical census \citep{Penny2019, Johnson2020} to studying entirely new, unknown populations \citep{DeRocco2023, Pruett2024, Fardeen2024}.

Importantly, astrometric microlensing allows for a direct \review{measurement of the angular Einstein radius of the lensing system} \citep{Hog1995, Walker1995, Miyamoto1995, DominikSahu2000} -- \review{which in some scenarios leads to a direct lens mass measurement},  a feat that has so far been achieved only for a handful of events with carefully planned follow-up (\citealt{Sahu2017}, \citealt{Zurlo2018}, \citealt{Sahu2022} and \citealt{Lam2022, Lam2022_sup}, \citealt{McGill2023}), but will become possible using routine GBTDS observations in the \romanst~era \citep[e.g.,][]{SajadianSahu2023}. Without input from astrometric measurements, mass estimates for dark lenses have usually been reliant on additional information from Galactic model priors and burdened with large uncertainties \citep[e.g.,][]{Wyrz2016, Howil2025, Kaczmarek2025}. Furthermore, astrometric measurements can resolve degeneracies that lead to multiple solutions in photometric-only models, offering a definitive determination of the proper motions of both the source and the lens \citep[e.g.,][]{Rybicki2018, Kaczmarek2022}, which is particularly useful for studying natal kicks. The potential of microlensing surveys in constraining kicks has already been signaled by \citet{Koshimoto2024}, who analysed the impact of black hole kick distributions on microlensing event rates, concluding that the detection of OB110462 favours low-kick scenarios.

In anticipation for the deluge of data expected from \romanst, we now conduct detailed simulations of \romanst-like event yields and detectability, focusing specifically on neutron star lenses and their measurable parameters. The paper is structured as follows. In Section \ref{sec:observables}, we introduce microlensing observables and equations used throughout this work. In Section \ref{sec:data} we describe the creation of mock datasets used in this work, including the underlying Galactic models as well as expected characteristics (cadence, precision, etc.) and detectability criteria for the GBTDS. In Section \ref{sec:results} we present the simulation results. We find a region in the observable parameter space that is preferentially occupied by NS lenses. We demonstrate that with \romanst~data, relatively high-purity samples of isolated NS can be constructed, which has not been possible with previous surveys. Yields of $\approx3-4 \cdot 10^3$ detectable NS and $\approx3 \cdot 10^4$ detectable stellar remnants overall are expected. As those events have been selected for both photometric and astrometric signal, their masses can be directly measured by $\romanst$, providing an invaluable resource to constrain remnant mass distributions \review{and the underlying physics}. We outline recommendations for classification of NS lenses and application of upcoming \romanst~results. Finally, in Section \ref{sec:discussion} we discuss and summarise our results.

\section{Photometric and astrometric microlensing observables}
\label{sec:observables}

The on-sky angular scale of a microlensing event caused by a lens of mass $M_L$ at a distance $D_L$, deflecting light from a source located at a distance $D_S$, is defined by the angular Einstein radius $\theta_{\rm E}$:
\begin{equation}
    \theta_{\rm E} = \sqrt{\frac{4GM_L}{c^2} \left(\frac{1}{D_L} - \frac{1}{D_S}\right)}
    \label{eq:thetaE}
\end{equation}

The timescale of the event is defined as the Einstein time $t_{\rm E} = \theta_{\rm E}/\mu_{\rm rel}$, i.e. the time needed to cross the angular Einstein radius in the relative lens-source motion $\mu_{\rm rel}$.

For a lens-source separation of \vec{u}, in units of $\theta_{\rm E}$, the source is amplified by a factor of $A$ \citep{Paczynski1996}:

\begin{equation}
    A = \frac{|\vec{u}|^2 + 2}{|\vec{u}|\sqrt{|\vec{u}|^2 + 4}}; \Delta m = -2.5\log_{10}(A)
    \label{eq:ampl}
\end{equation}

The only parameters tied to physical properties of the lens possible to constrain from the lightcurve alone are $t_{\rm E}$ and the microlensing parallax $\vec{\pi_{\rm E}}$. \review{Only $t_{\rm E}$ is routinely available, as parallax deviation is a second-order effect that is difficult to constrain and not commonly measured. The microlensing parallax $\vec{\pi_{\rm E}}$ is a vector with magnitude of:}

\begin{equation}
    \pi_{\rm E} = \frac{1}{\theta_{\rm E}} \left( \frac{1 \textnormal{AU}}{D_L} - \frac{1 \textnormal{AU}}{D_S} \right)
\end{equation}
\review{and the direction of relative lens-source motion \citep[e.g.,][]{Gould2004}.} Microlensing parallax can be constrained by modelling the imprint of the observer's annual motion around the Sun, which causes deviations from straight-line apparent lens-source motion, on the lightcurve. The relative position of the lens with respect to the source, in units of $\theta_{\rm E}$, is modelled as:

\begin{equation}
    \vec{u}(t) = \vec{u_0} + \frac{t - t_0}{t_{\rm E}} \vec{\hat{\mu}_{rel}} + \vec{\pi}(\vec{\pi_{\rm E}}, t)
    \label{eq:u}
\end{equation}

where $\vec{u_0}$ is the closest approach perpendicular to the relative motion and $\vec{\pi}$ denotes parallax deviations from straight-line motion caused by projected orbital movement of the observer.

In addition to photometric signal, the source appears deflected away from the lens, causing an apparent astrometric shift $\vec{\delta}(\vec{u})$ \citep{Hog1995, Walker1995, Miyamoto1995}.
\begin{equation}
    \vec{\delta}(\vec{u}) = \frac{\vec{u}}{|\vec{u}|^2 + 2} \theta_{\rm E}
    \label{eq:delta}
\end{equation}

Therefore -- in addition to photometric observables ($t_{\rm E}$, $\vec{\pi_{\rm E}}$) discussed above -- astrometric microlensing allows for the direct measurement of $\theta_{\rm E}$ and a complete resolution of the event, including measuring the lens mass $M_L$.

\section{Data}
\label{sec:data}

\subsection{Simulation procedure}

\begin{table*}
\begin{center}
\begin{tabular}{p{90mm}p{88mm}}
    \hline\hline 
    Parameter & Value \\\hline
    Milky Way escape velocity & $550$kms$^{-1}$ \citep{Piffl2014} \\
    Sun-Galactic center distance & $8.3$kpc \\
    Mean BH natal kick & $100$kms$^{-1}$ \\
    Mean NS natal kick & \{$150$kms$^{-1}$, $250$kms$^{-1}$, $350$kms$^{-1}$, $450$kms$^{-1}$\} \\
    Initial-Final Mass Relation (IFMR) & SukhboldN20 \citep{Sukhbold2014, Sukhbold2016, Woosley2017, Woosley2020}\\
    Extinction law & \citet{Damineli2016} \\
    Bar dimensions (radius, major axis, minor axis, height) & $(2.54, 0.70, 0.424, 0.424)$ kpc \\
    Bar angle (Sun–Galactic center, 2nd, 3rd) & $(62.0, 3.5, 91.3)$ $^{\circ}$ \\
    Bulge velocity dispersion (radial, azimuthal, z) & $(100, 100, 100)$ kms$^{-1}$ \\
    Bar pattern speed & 40.00kms$^{-1}$ kpc$^{-1}$ \\
    Multiplicity & Singles \\
    \hline
\end{tabular}
\end{center}
\caption{\label{tab:popsycle_params} \texttt{PopSyCLE} simulation parameters used in this work. The implementation of the IFMR is described in detail in \cite{Rose2022}; Galactic parameters are consistent with ''v3'' in \citet[App.~A]{Lam2020}. We run four simulations differing by NS natal kick distributions.}
\end{table*}

\begin{figure}
    \centering
    \includegraphics[width=\columnwidth, trim=0cm 0cm 0cm 0cm]{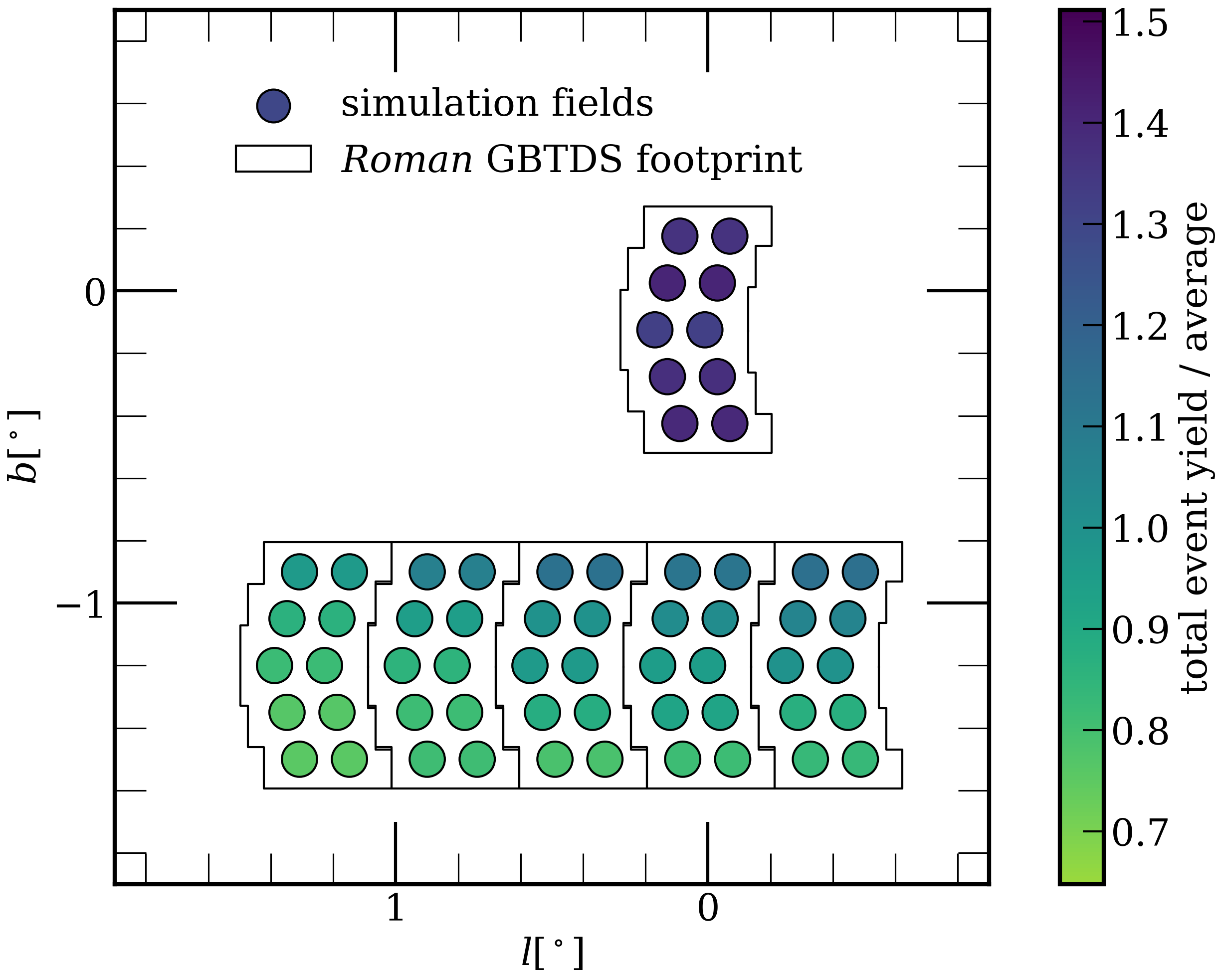}
    \caption{Layout of the simulation fields, overplotted on the \romanst~GBTDS footprint as recommended by the Roman Observations Time Allocation Committee. Colours represent the total number of simulated microlensing events per simulation field over the average for all fields in the run with the {\tt PopSyCLE} \citep{Lam2020} default 350 km/s mean NS kick velocity (before applying the detectability cuts).}
    \label{fig:footprint}
\end{figure}

To perform our simulations, we use the state-of-the-art software for simulating microlensing events in the Milky Way, \texttt{PopSyCLE} \citep[Population Synthesis for Compact-object Lensing Events;][]{Lam2020}. \texttt{PopSyCLE} uses the Besançon model \citep{Robin2004} implemented in \texttt{Galaxia} \citep{Sharma2011} to generate a synthetic stellar survey within a given circular field. Stellar remnants are generated and injected into the survey area by evolving clusters using \texttt{SPISEA} \citep[Stellar Population Interface for Stellar Evolution and Atmospheres;][]{Hosek2020}, a software package generating single-age, single-metallicity populations. SPISEA includes initial mass functions, multiplicity distributions, metallicity-dependent stellar evolution and atmosphere grids, and extinction laws; in order to accommodate mock microlensing survey generation, it has been extended to also include initial-final mass relations (IFMRs) between progenitors and their remnants \citep{Lam2020}. Following \citet{Kaczmarek2025, Abrams2025}, we choose to use the SukhboldN20 IFMR, which is based on a suite of simulations by \citet{Sukhbold2014, Sukhbold2016, Woosley2017, Woosley2020} and is unique in its use of both metallicity dependence and inputs from recent supernova explosion models. The simulated events belong to four lens classes, which hereafter we denote as $\text{class}_L$: Star, WD (white dwarf), NS (neutron star) and BH (black hole). Newly generated BH and NS receive a natal kick in a random direction, drawn from a Maxwellian distribution with a specified mean. In this work, we vary the NS kick distributions to assess their impact on microlensing observables. All input parameters of the \texttt{PopSyCLE} simulations can be found in Table \ref{tab:popsycle_params}. We choose synthetic survey fields to approximate the \romanst~footprint, which we plot using the {\tt gbtds\_optimizer} software\footnote{\url{https://github.com/mtpenny/gbtds_optimizer}}. We present the field layout in Figure \ref{fig:footprint}.

\subsection{Simulated event datasets}

Lensing events are picked out of the simulated Galactic populations by selecting lens--luminous source pairs with a minimum separation of < $2\theta_{\rm E}$ within the simulated survey time of 5.0 years, matching the scheduled 10-season span of GBTDS. With this procedure, we create four datasets of lensing events. Star, WD and BH populations are the same in each dataset, as the simulations were initiated with the same seed; the NS populations show systematic shifts in parameter distributions with varying kick velocity.

We make all simulated microlensing event datasets generated in this work publicly available within \texttt{popclass}\footnote{\url{https://github.com/LLNL/popclass}}, a software package for probabilistic classification of microlensing events \citep{popclass}.

\subsection{Post-processing for detectability with \romanst}
\label{sec:postproc}

To select only events that can be fully characterized with GBTDS photometry and astrometry, we apply cuts similar to \citet{Fardeen2024}:

\begin{itemize}
    \item already at the stage of executing \texttt{PopSyCLE}, we have applied a $|u_0| < 2$ cut, corresponding to photometric signal of $\Delta m = 0.06$ at peak. \\
    
    \item we convert the \texttt{PopSyCLE} apparent ($H$, $J$, $K$) magnitudes to $F146$ \citep[using Eq. (1) from][]{Wilson2023} and apply a $F146 < 23$ cut. In case of star lenses, this cut is applied to the blended lens+source flux. Using image-level simulations, \citet{Wilson2023} have determined $F146 = 22$ to be the transition at which the background's contribution to noise starts exceeding that of the source. We estimate that, with tailored data processing techniques, it should be possible to achieve a drop of $\approx$1 magnitude below this boundary to obtain scientifically useful images of the source. This is a rough estimate, as it is difficult to predict future mission performance; it should be re-evaluated after \romanst's start of operations.\\
    
    \item we create a set of observing times $\mathcal{T}_{GBTDS}$, which follows the most up-to-date recommendations for the GBTDS survey design, i.e. 6 high-cadence seasons of 70.5 days, centered around the vernal and autumnal equinoxes and filled with 12.1-minute cadence observations, separated in half by a gap of 4 low-cadence seasons. We also include gap-filling observations every 3 days in the 4 low-cadence seasons. (This set is the same for all fields, as they will have the same cadence; the choice of a starting point is arbitrary and does not influence observed event distributions.)\\
    
    \item we (conservatively) assume that to fully characterize the event, we require the closest approach to happen within the bounds of the survey time: $\min(\mathcal{T}_{GBTDS}) < t_0 < \max(\mathcal{T}_{GBTDS})$.\\
    
    \item to select only events with detectable astrometric signal, we require $\Delta_{ast} > \sigma_{ast}/\sqrt{N}$, where $\Delta_{ast} = \max |\vec{\delta}(t_i) - \vec{\delta}(t_j)|: t_i, t_j \in \mathcal{T}_{GBTDS}$. We assume $N = 119$ for stacking observations taken within 24 hours. We use Eqs.~\ref{eq:u},~\ref{eq:delta} to calculate $\delta(t): t \in \mathcal{T}_{GBTDS}$ values (in Eq.~\ref{eq:u}, we assume straight-line motion as parallax deviations are typically very small and, with event parameters $t_0, u_0$ distributed randomly, do not have significant impact on population distributions); in case of star lenses, we correct $\Delta_{ast}$ for blending with the lens. We follow Eq. (14) from \citet{Fardeen2024} to obtain $\sigma_{ast}$ [mas] for a single observation as a function of magnitude; this formula is based on a straight-line fit to \romanst~simulations from \citet{SajadianSahu2023}.\\
    
    \item analogously, to select only events with detectable photometric signal, we require $\Delta_{phot} > \sigma_{phot}/\sqrt{N}$, where $\Delta_{phot} = \max |\Delta m(t_i) - \Delta m(t_j)|: t_i, t_j \in \mathcal{T}_{GBTDS}$. We use Eqs.~\ref{eq:ampl},~\ref{eq:u} to calculate $\Delta m(t): t \in \mathcal{T}_{GBTDS}$ values; in case of star lenses, we correct $\Delta_{phot}$ for blending with the lens. To obtain $\sigma_{phot}$ for a single observation, we use the relative flux error model depicted in Figure 4 from \citet{Wilson2023} image simulations and approximate it with a polynomial fit:

    \begin{equation}
    \begin{aligned}
        \log_{10}\left(\sigma_{F}/F\right)= & -1.379 \cdot 10^{-4} m^3 
        + 2.817 \cdot 10^{-2} m^2 \\
        & - 7.389 \cdot 10^{-1} m + 5.319 \ \textnormal{[ppt],}
    \end{aligned}
    \end{equation}

    where $m$ is the $F146$ magnitude, and convert it to magnitudes:

    \begin{equation}
    \begin{aligned}
        \sigma_{phot} = \frac{2.5}{\ln 10}\frac{\sigma_{F}}{F}\textnormal{.}
    \end{aligned}
    \end{equation}
    
\end{itemize}

After applying all cuts, we retain 2.8\% of all simulated events, including 4.8-6.1\% (dependent on kick velocity) of all simulated events with NS lenses. We note that after correcting for blending with luminous lenses, we actually retain slightly more events (though some previously flagged as detectable are lost due to decreased $\Delta_{phot}$ and $\Delta_{ast}$, more new lenses are included due to passing the magnitude cut with increased joint flux). Hereafter, we use 'detectable' to describe events passing all cuts, i.e. both photometrically and astrometrically detectable, unless indicated otherwise. \review{We make datasets containing parameters of these events publicly available in the built-in model library of the \texttt{popclass} classification software\footnote{\url{https://github.com/LLNL/popclass}} as {\tt gbtds\_nskick150, ..., gbtds\_nskick450}. We also include the full {\tt PopSyCLE} output for the simulated events in an auxiliary dataset, which will be made publicly available upon the acceptance of this paper. In this dataset, we include both the event samples before and after the pre-processing step, in case the users choose to apply their own detectability criteria} (e.g. if \romanst~performance allows for relaxing some cuts in the future). 

\review{In this study, we do not attempt to assess the constraints that the GBTDS data will enable on $t_{\rm E}, \theta_{\rm E}, \pi_{\rm E}$ for a given detectable event, and we leave this to future work. Implicitly assumed in the work that follows is that we can measure $t_{\rm E}$ and $\theta_{\rm E}$ precisely. This assumption ends up being reasonable for the set of events we focus on. The `spur' lenses are located at the high-$\theta_{\rm E}$ end of the parameter space and have $t_{\rm E} \sim 10$ days, which are likely to be well-constrained from the $\approx12$-minute cadence photometric and astrometric observations of the GBTDS. This point has been documented in several studies. \citet{Kaczmarek2022} find $< 1\%$ constraints on both $t_{\rm E}$ and $\theta_{\rm E}$ with modelling of simulated GBTDS~observations of a nearby NS lensing event. Similarly, \citet{Lu2025} find $< 1\%$ constraints on $t_{\rm E}$ and $< 10\%$ on $\theta_{\rm E}$ with modelling of simulated GBTDS~observations for an example BH lens. \citet{Terry2025} find $\approx90\%$  of their 3000 simulated GBTDS lenses (chosen to be a representative population of planetary hosts, i.e. systematically lower-mass than stellar remnants) have a constraint on $t_{\rm E}$ of $\leq10\%$ derived from the light curve alone. On the astrometric front, \citet{SajadianSahu2023} find that >99\% of their simulated BH events observed by the GBTDS have a constraint on $\theta_{\rm E}$ of $\leq10\%$ with the majority of them also having a constraint on $t_{\rm E} \leq 10\%$, regardless of the mass function assumed. Finally, \cite{Fardeen2024} find that many wide-separation ($u_{0}>2$) events causes by lenses with masses around one solar mass have constraints on $\theta_{E}$ and $t_{E}$ of $<10$\% from astrometry alone.}

\begin{figure}
    \centering
    \includegraphics[width=0.95\columnwidth, trim=0cm 0cm 0cm 0cm]{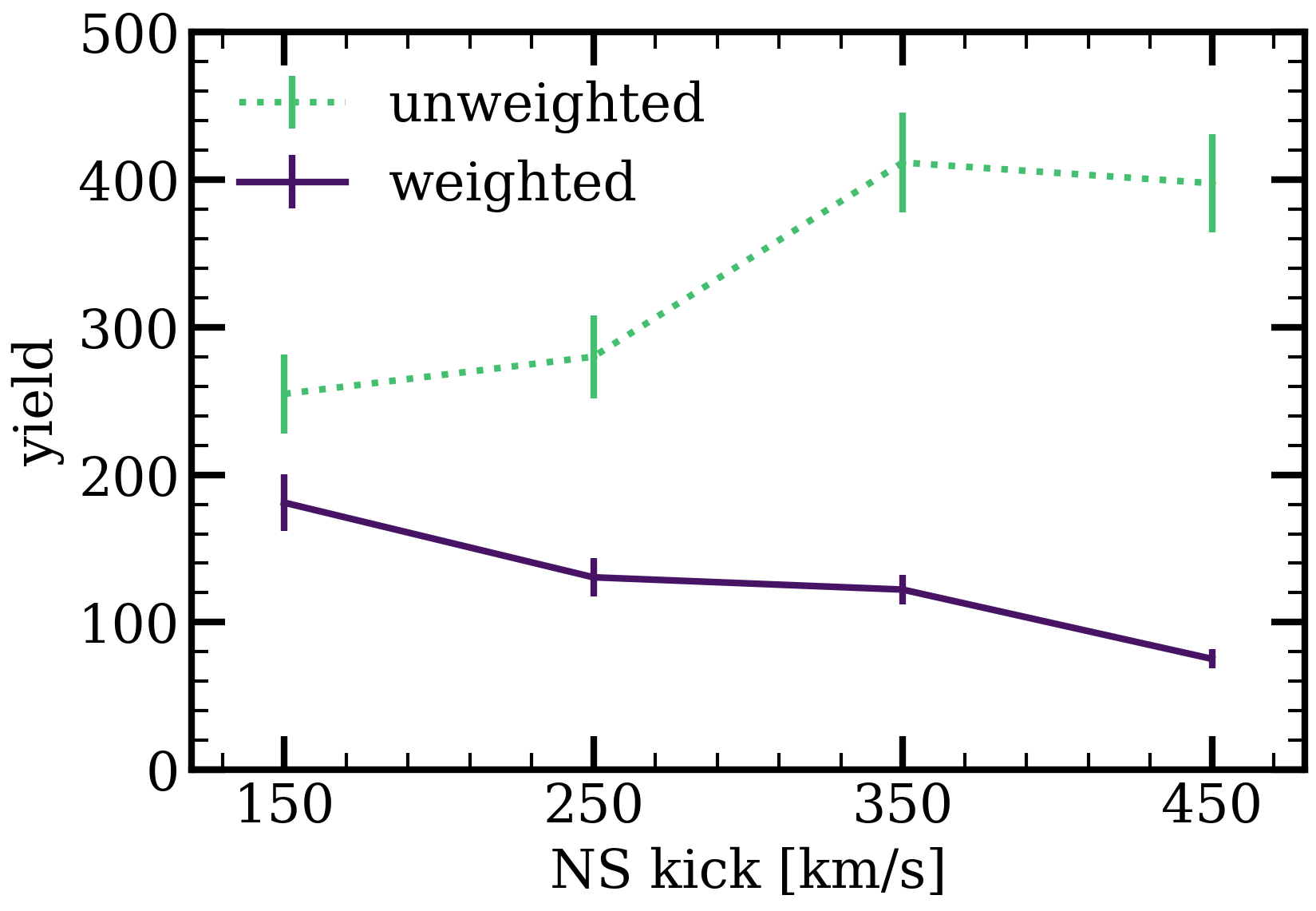}
    \caption{\highlight{Yield of events with NS lenses detectable with \romanst~in both photometry and astrometry over the entire GBTDS duration as a function of mean kick velocity $v_{\rm kick}$, before (green, dotted) and after (purple, solid) applying the effect of natal kicks on volume density of NS in the Galaxy. Errorbars represent Poisson noise; the $\sigma^2 = \sum_i w_i^2$ formula \citep{Barlow1987} is applied to weighted Poisson counts. While larger kicks increase the probability of lensing by a single NS, this effect is counterbalanced in weighting by a lower amount of NS available as lenses in the inner Galactic regions.}}
    \label{fig:ns_yields}
\end{figure}

\review{In summary, lens parameter constraints from GBTDS observations are a topic of ongoing study, but $\theta_{\rm E}$ and $t_{\rm E}$ for stellar remnants events are generally expected to be well-constrained. We note a key caveat is that these simulations and predictions generally do not consider potential contaminants. Other effects including astrometric binaries \citep[e.g.,][]{Jankovic2025} or binary source/binary lens microlensing \citep[e.g.,][]{Abrams2025} can be confused for point source -- point lens events, though the extent to which this will impact simultaneous photometric and astrometric observations of GBTDS is yet unclear. Assuming no contamination makes the following work an optimistic upper bound in this respect.}

\section{Results}
\label{sec:results}

\subsection{Detectable event yields}

To calculate the detectable event yield expected from GBTDS, we upscale the retained event counts from Section~\ref{sec:postproc} by $S_{GBTDS}/S_{sim}$, the ratio of the full GBTDS footprint area to the total area of simulation fields (depicted in Figure~\ref{fig:footprint}). We assume Poisson noise on retained event counts. We note that all reported errorbars on event yields only reflect Poisson noise of the simulation run and not any uncertainties in the Galactic parameters of the underlying simulation. We estimate GBTDS will include \highlight{$\approx$11000 detectable events overall, including 8710 $\pm$ 160 stars and 2229 $\pm$ 79 white dwarfs.}

\highlight{Our treatment of massive remnants is more complex. With out-of-the-box simulation results, we retain 168 $\pm$ 22 detectable BH events and (254 $\pm$ 27, 280 $\pm$ 28, 412 $\pm$ 34 and 398 $\pm$ 33) detectable NS events for $v_{\rm kick}$ of (150, 250, 350, 450) km/s, respectively. However, those yields are likely to be overestimated. \citet{Sweeney2022} have studied the expected Galactic distributions of stellar remnants, finding neutron stars and black holes to have more diffuse spatial distributions than the visible Galaxy, primarily due to natal kicks changing their Galactic orbits. As {\tt PopSyCLE} currently does not include this effect, keeping remnants at fixed Galactic positions, the yields we obtain in the simulations described are effectively upper limits (i.e. assuming all NS are young and have not yet had time to migrate outwards).}

\begin{figure*}
   \centering
   \includegraphics[width=2\columnwidth, trim=0cm 0cm 0cm 0cm]{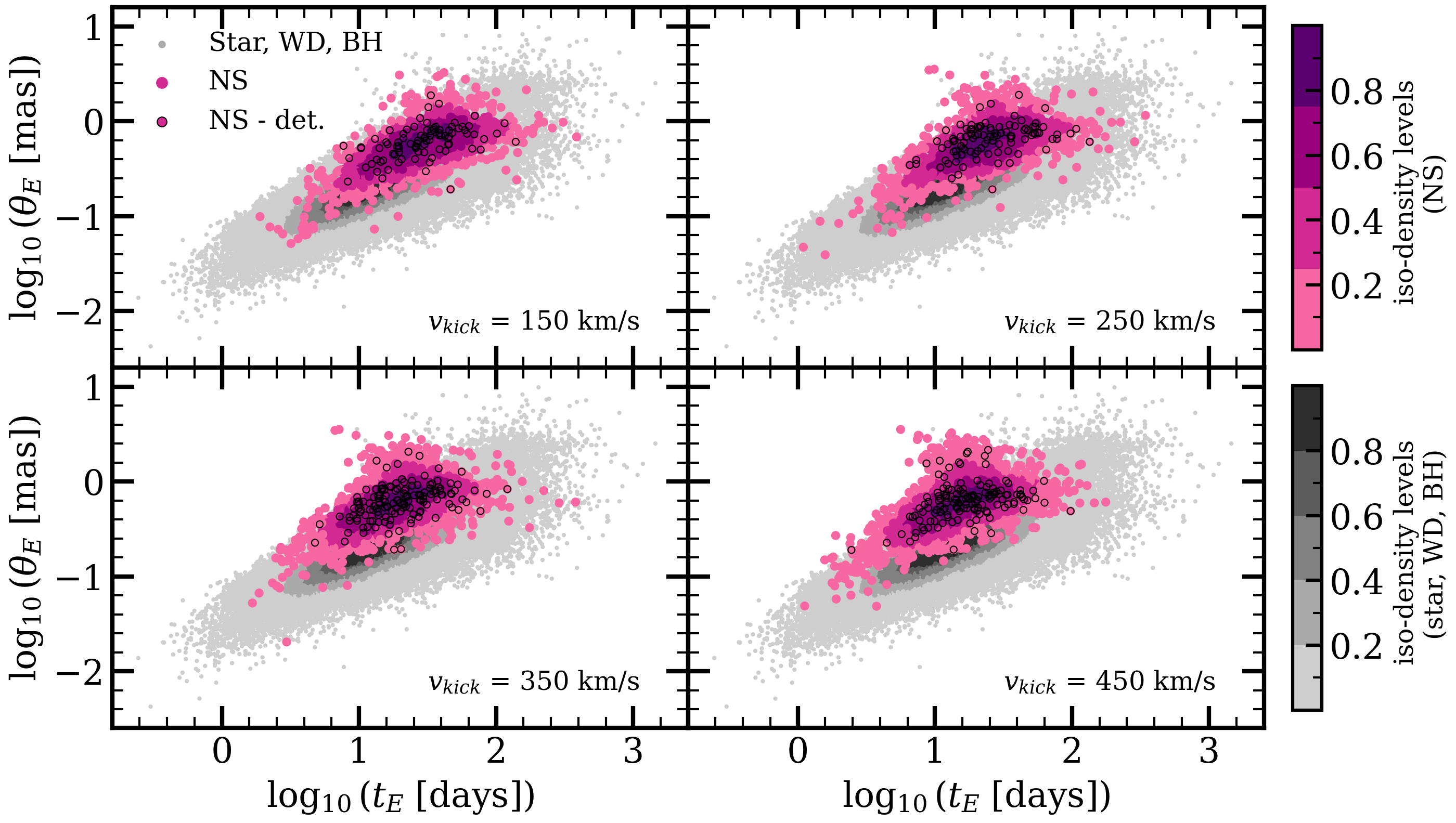}
   \caption{Distributions of NS (purple) and non-NS (grey) events in $\log_{10} t_{\rm E}$--$\log_{10}\theta_{\rm E}$ space. All simulated events regardless of passing detectability cuts are plotted; shade denotes iso-density levels estimated with a Gaussian KDE. Black circles highlight the NS events that passed the detectability cuts. The NS distribution exhibits a characteristic `spur' feature that becomes stronger and more shifted leftwards with increasing $v_{\rm kick}$, yielding more detectable NS outlying from the main distribution.}
   \label{fig:ns_spur}
\end{figure*}

\highlight{We aim to provide more realistic yields by applying a re-weighting to our simulated events. We use a sample of 10$^5$ simulated NS from \citet{Sweeney2022} to model the pre-kick NS distribution. We use the {\tt StellarMortis} software\footnote{\url{https://github.com/David-Sweeney/StellarMortis}} \review{\citep{software:StellarMortis}} that assigns kicks to remnants and propagates them in the Milky Way potential to a present-day distribution. We modify {\tt StellarMortis} to include flexible Maxwellian kick distributions for a given $\bar{v}$ and generate four samples of 10$^5$ post-kick NS matching our four simulated kick distributions. We then bin these samples in Galactocentric $r$ and $|z|$ (assuming the distributions are axisymmetric and symmetric in $z$) to minimize random noise. To each simulated NS lens $i$ located at $r_i$, $z_i$, we assign a weight $w_i = N_{\rm post}(r_i, |z_i|) / N_{\rm pre}(r_i, |z_i|)$ -- equal to the ratio of post-kick to pre-kick NS counts in a given bin. Then, our density-corrected yields are equal to the sum $\sum_i w_i \cdot S_{GBTDS}/S_{sim}$ over all NS lenses in the simulation. We visualize the NS yields as a function of $v_{\rm kick}$ in Figure~\ref{fig:ns_yields} for both the weighted and unweighted case. The corrected yields are (181 $\pm$ 19, 131 $\pm$ 13, 122 $\pm$ 10, 75 $\pm$ 6) respectively with increasing  $v_{\rm kick}$. The weighting and its limitations are discussed in more detail in Section \ref{sec:discussion}.}

\highlight{The black hole kick distribution is very uncertain. BH are usually expected to have a similar kick momentum distribution to NS, therefore lower velocities \cite[e.g.][]{Fryer2001}. However, \citet{Repetto2012} found evidence for BH kick \textit{velocities} similar to those of NS, and \citet{Janka2013} offered a theoretical mechanism that could explain them. \citet{Nagarajan2025} find evidence for strong kicks for some BH and weak to none for others with Gaia DR3 kinematics. We note several recent works on newly discovered BH support the low-kick hypothesis \citep[e.g.,][]{Kotko2024, Koshimoto2024, Burdge2024}. Here, for consistency with the {\tt PopSyCLE} default settings used in our simulations, we assume $v_{\rm kick}=100$ km/s for black holes to apply the same yield correction as for NS lenses. As there are only 75235 black holes in the \citet{Sweeney2022} remnant sample, we use the full sample to generate weights, resulting in a corrected yield of 152 $\pm$ 21 BH lens events.}

As the recommended GBTDS strategy is not final yet, we also estimate detectable event yield assuming no gap-filling (low-cadence) observations. We find that without these observations, we retain only 62\% of detectable events (60\% for star lenses, 66\% for white dwarf lenses and 91\% for black hole lenses). Dependent on $v_{\rm kick}$, we also retain (72, 66, 62, 55)\% of NS events (for $v_{\rm kick}$ of (150, 250, 350, 450) km/s respectively)\footnote{Including weighting for NS and BH; however, weighting has minimal (<1\%) impact on the percentages.}. Removing the gap-filling observations impacts remnant detections (and especially black hole detections) less, as populations of high-mass lenses have longer typical timescales. Strongly kicked neutron stars are the exception, as high relative lens-source motions cancel out that effect. This experiment proves the effectiveness of gap-filling observations in GBTDS, which can increase the detectable event yield by $\gtrsim$50\% at a relatively low observing time cost (only 1 hour of observations per 3 days outside the high-cadence seasons\footnote{Roman Observations Time Allocation Committee Final Report and Recommendations}).

\begin{figure*}
    \centering
    \includegraphics[width=2
    \columnwidth, trim=0cm 0cm 0cm 0cm]{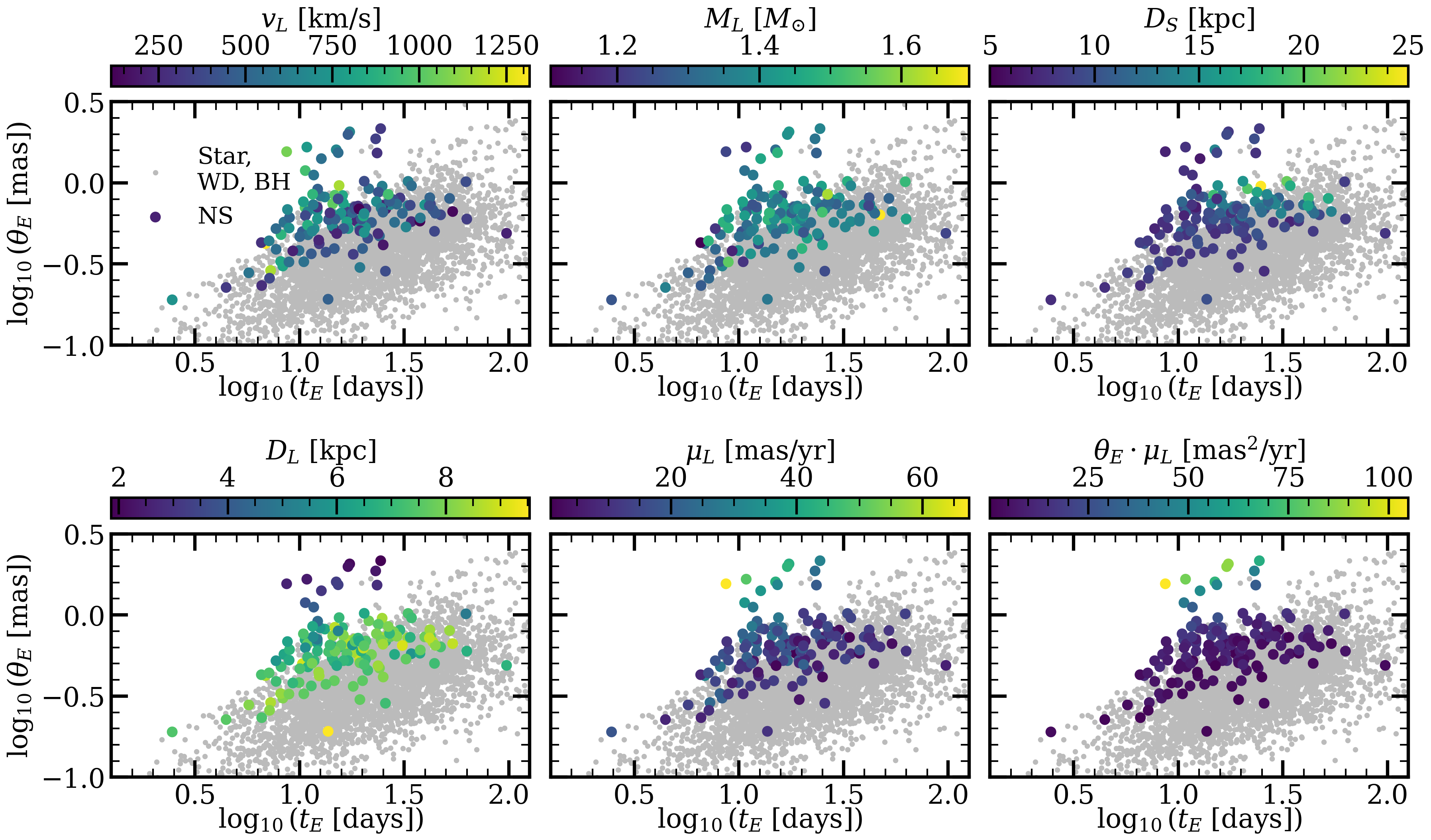}
    \caption{All six subplots present the same $\log_{10} t_{\rm E}$--$\log_{10}\theta_{\rm E}$ space distribution of detectable NS lenses from the $v_{\rm kick}$ = 450 km/s simulation run (coloured) compared to that of lenses from other (Star, WD, BH) classes (light grey). The NS lenses are coloured by event parameters: lens velocity $v_L$, lens mass $M_L$, source distance $D_L$, lens distance $\mu_L$, and a measure of background area subject to light deflection by the lens per unit time $\theta_{\rm E} \cdot\mu_{L}$, respectively (from top left row-wise). NS located in the `spur' region distinguishable from other lens classes are nearby, have high proper motions, and are relatively very likely to cause lensing events.}
    \label{fig:ns_spur_coloured}
\end{figure*}

\subsection{Distributions in $\log_{10} t_{\rm E} - \log_{10} \theta_{\rm E}$ space}

With simultaneous photometric and astrometric observations, we can measure three microlensing observables related to lens mass: $t_{\rm E}, \theta_{\rm E}, \pi_{\rm E}$. Of these, $t_{\rm E}$ and $\theta_{\rm E}$ are optimal for remnant classification, as both parameters can be well constrained from observations and increase with lens mass; on the contrary, $\pi_{\rm E}$ is difficult to measure precisely and decreases with lens mass. In photometry, while $t_{\rm E}$ is usually well-constrained from the lightcurve, $\pi_{\rm E}$ is subject to various degeneracies \citep{Smith2003, Gould2004} and notoriously hard to constrain, especially for remnant candidates \citep[e.g.][]{Kaczmarek2022, Golovich2022}. In astrometry, for massive remnants (NS, BH) the $\pi_{\rm E}$ contribution is 1-2 orders of magnitude smaller than the overall astrometric deviation with which we measure $\theta_{\rm E}$. We operate in $\log$--$\log$ space as it is optimal for separation of lens classes \citep[e.g.,][]{Lam2020, Fardeen2024, Kaczmarek2025}.

With increasing $v_{\rm kick}$, we find increasing representation of NS that are outliers from the main distribution in the $\log_{10} t_{\rm E}$--$\log_{10}\theta_{\rm E}$ space. In Figure~\ref{fig:ns_spur}, we plot the simulated event populations in this space, including all NS and non-NS events regardless of cuts. We show these outliers are sampled from a continuous `spur' structure, which is characteristic to the NS population and diagonally offset from the main $\log_{10} t_{\rm E}$--$\log_{10}\theta_{\rm E}$ distribution. The region covered by the `spur' has a high ratio of NS to total event numbers \review{(for a more detailed discussion and visualization of the number of NS events and non-NS contaminants expected in this structure, see Figs.~\ref{fig:ns_spur_coloured_weighted} and \ref{fig:purity_recall} and the corresponding descriptions). We} propose utilising this feature for constructing optimal NS candidate samples. 

\subsection{Properties of detectable NS events}

\highlight{We now look into how the physical parameters of NS} relate to their positions in that space, and in particular to their membership in the `spur' feature. In Fig.~\ref{fig:ns_spur_coloured}, we present distributions of several event parameters across the $t_{\rm E} - \theta_{\rm E}$ space for $v_{\rm kick}$ = 450 km/s, the simulation in which the spur is most prominent.

\begin{figure}
    \centering
    \includegraphics[width=0.9
    \columnwidth, trim=0cm 0cm 0cm 0cm]{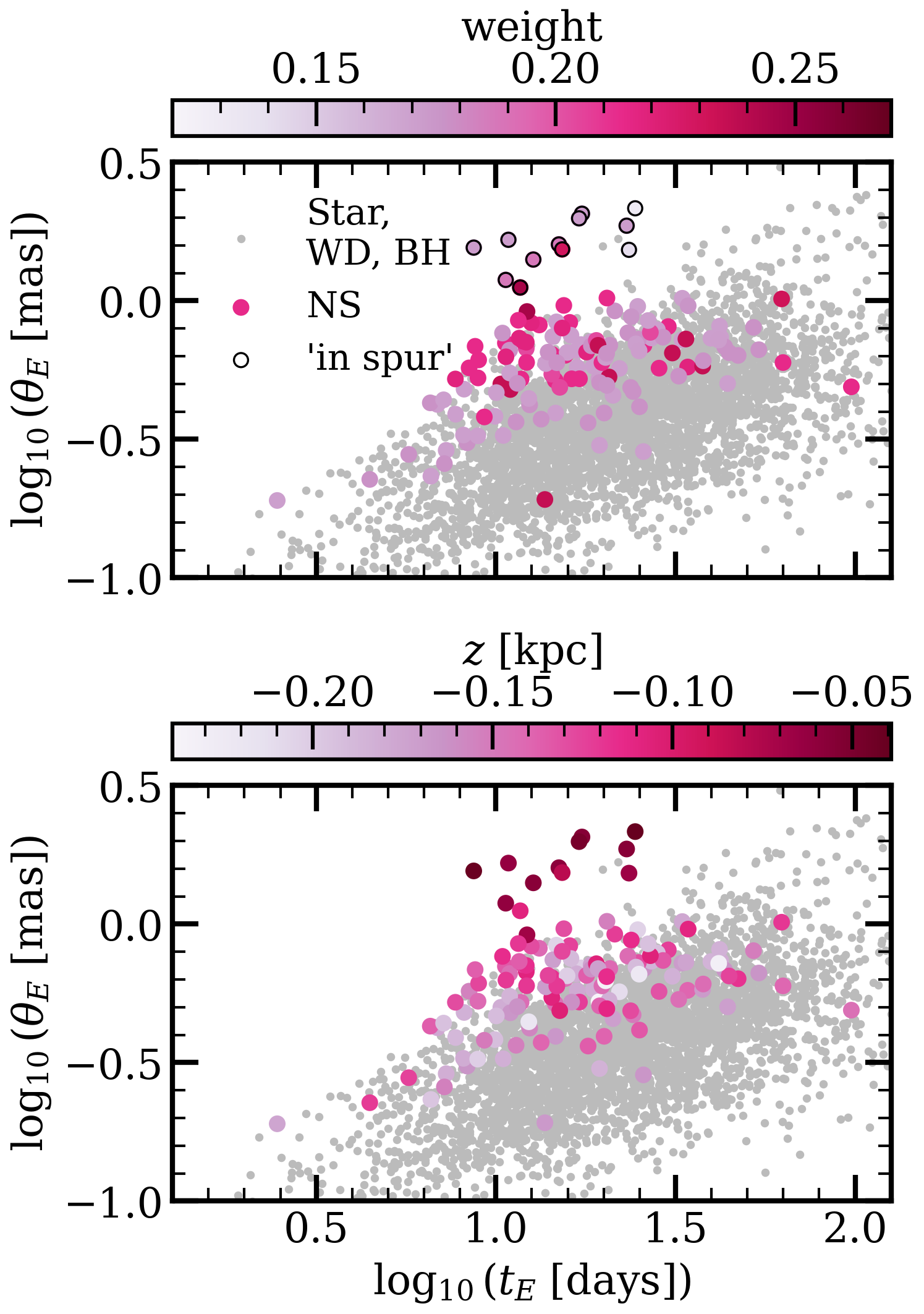}
    \caption{\highlight{Like Fig.~\ref{fig:ns_spur_coloured}, but coloured by parameters relevant to correcting for changed NS density: weight and Galactocentric $z$ coordinate. In the top subplot, black outlines denote events we select as `in spur' to estimate the total yield of $\approx$6 expected `spur' NS lenses in GBTDS, assuming $v_{\rm kick} = 450$ km/s.}}
    \label{fig:ns_spur_coloured_weighted}
\end{figure}

While weak trends in $v_L$ and $M_L$ are present (NS with higher velocities and lower masses on average cause shorter events), neither of these parameters is responsible for the spur. We also note there is a region preferentially occupied by NS events with distant sources at approximately $\log_{10} t_{\rm E}=1.5$, $\log_{10}\theta_{\rm E}=0$, which is also unrelated to the spur and not distinguishable from other event classes.

Conversely, in the bottom panels, we present lens parameters correlated with membership in the spur. The NS lenses located in the spur are much more nearby and have higher proper motions. We conclude the spur contains \textit{mesolenses} - single objects that are naturally high-probability lenses, even though the total contribution to the lensing optical depth from their population may not be very large \citep{DiStefano2008b, DiStefano2008a}. \citet{DiStefano2008a} states that mesolenses are caused by a combination of a large Einstein ring, fast angular motion, and a dense background field. For simplicity, let us assume constant source density within a simulation field, random background source motion, and constant $\theta_{\rm E}$ per lens. Then, contribution to event rate $\Gamma$ in the field from a single object is $\sim \theta_{\rm E} \cdot\mu_{L}$, i.e. proportional to the area swept by the Einstein ring per unit time. We plot $\theta_{\rm E} \cdot\mu_{L}$ in the bottom right panel, showing NS lenses located in the spur vastly exceed other NS lenses in contribution to event rate, i.e. are very likely to cause lensing events.

Such objects are of great interest as they can cause repeated events \citep[e.g.][]{Bramich2018}, allowing high-precision mass determination with several independent mass measurements. To maximise mass measurement precision, mesolens candidates should be vetted for possible repeated events. For high-$\theta_{\rm E}$ events with \romanst -like observations, lens motion can be precisely constrained even if the lens is not seen \citep[e.g.][]{Kaczmarek2022}. If motions of found mesolenses can be constrained well enough from \romanst~data, predictions for future events should also be made to enable follow-up observations.

\highlight{Fig.~\ref{fig:ns_spur_coloured} shows the lenses as recovered from the simulation -- without the Galactic NS density correction. To visualize the effect of this correction, we also plot the weights and $z$ positions in a supplementary Fig.~\ref{fig:ns_spur_coloured_weighted}, which shows that the spur lenses have systematically lower weights. This result may seem counterintuitive, as the expansion of the NS distribution will result in lenses migrating from the Bulge to higher Galactocentric radii -- therefore, a higher fraction of nearby mesolenses could be expected. However, this is explained by the lower panel: as the line of sight is fixed, the nearby lenses falling into the \romanst~fields have $z \approx z_\odot \approx 0$. As the post-kick NS distributions have inflated scale heights, weights increase with $|z|$. Overall, for $v_{\rm kick} = 450$ km/s, we can expect roughly $\sum_i w_i \cdot S_{GBTDS}/S_{sim} \approx$ 6 NS lenses in the spur. Due to the small sample size, every detection of a lens in the spur region will be extremely valuable for population constraints. We suggest all lenses found in that region of $\log_{10} t_{\rm E}$--$\log_{10}\theta_{\rm E}$ space should be preferentially allocated follow-up resources (e.g. spectroscopy for source characterization).}

\begin{figure}
    \centering
    \includegraphics[width=\columnwidth, trim=0cm 0cm 0cm 0cm]{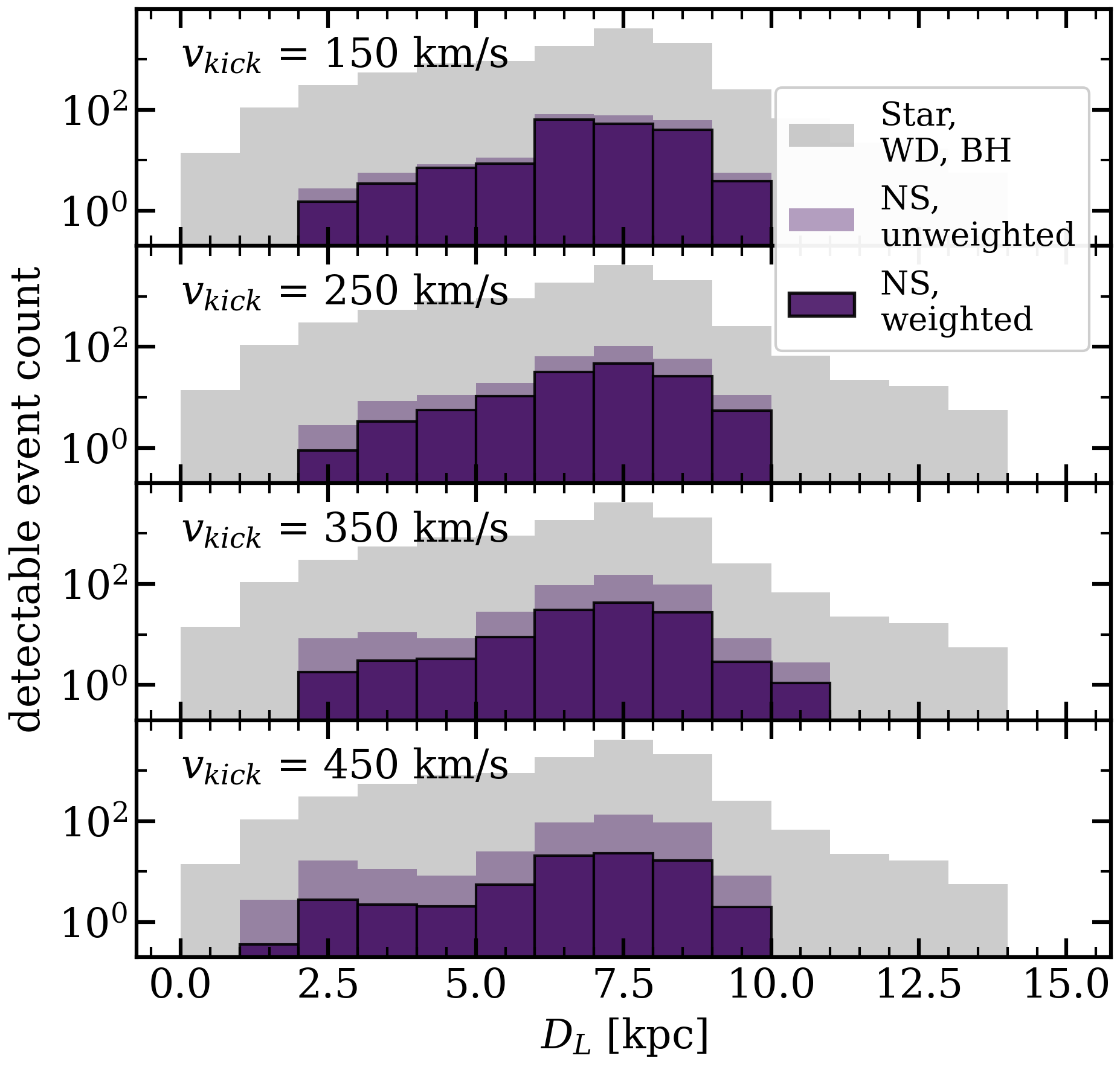}
    \caption{Distributions of detectable NS lens distances for all four runs, with $v_{\rm kick}$ increasing from top to bottom. \highlight{The unweighted distance distribution is plotted with light purple, while the weighted (with post-kick Galactic density correction) distance distribution is plotted with dark purple with black contours.} The distance distribution for all other detectable lenses ($\text{class}_L \in \{\text{Star, WD, BH}\}$) is plotted in light grey. \highlight{All yields are adjusted to the total survey area.}}
    \label{fig:dist_hist}
\end{figure}

To follow up on the correlation between $D_L$ and location in the spur, we present lens distance histograms for all simulations in Figure \ref{fig:dist_hist}. We find lens distance distribution is dependent on $v_{\rm kick}$, making it useful for NS kick constraints. \highlight{While the nearby mesolenses are, again, more affected by downweighting, high $v_{\rm kick}$ still yield flatter, more tailed lens distance distributions. As in the case of the spur structure, large kicks are responsible for effects both strengthening (generating nearby mesolenses) and weakening (preferential drop in density of nearby lenses along a fixed l.o.s. due to their low $|z|$) of a distinct structure in parameter distribution. Detailed studies on these counteracting effects will be necessary to leverage upcoming data for population constraints.}

\subsection{NS classification}

\begin{figure}
    \centering
    \includegraphics[width=\columnwidth, trim=0cm 0cm 0cm 0cm]{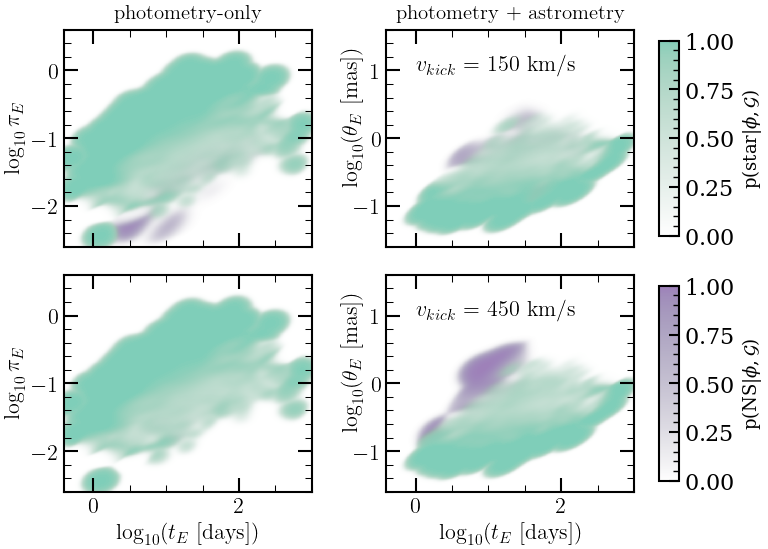}
    \caption{\textit{Left}: Relative class probabilities in $\log_{10} t_{\rm E}$--$\log_{10}\pi_{\rm E}$ -- the space of observables accessible from photometric observations only. The probability map is constructed using all photometrically detectable events. \textit{Right:} Relative class probabilities in $\log_{10} t_{\rm E}$--$\log_{10}\theta_{\rm E}$ -- the space of observables possible to measure with high precision with the combination of photometry and astrometry. The probability map is constructed using all photometrically and astrometrically detectable events. Each relative probability map is presented for $v_{\rm kick} = $ 150 km/s \textit{(top)} and 450 km/s \textit{(bottom)}. High $v_{\rm kick}$ results in a parameter distribution that is more distinguishable from other lens classes. In particular, in the $\log_{10} t_{\rm E}$--$\log_{10}\theta_{\rm E}$ space, there is a region preferentially occupied by NS, which reaches $p(\text{NS}| \phi, \mathcal{G}) \approx 1$ and could be used for constructing optimal samples of isolated NS candidates. (For clarity of the figure, the WD and BH lens class probabilities are not plotted, as the Star class is the main source of confusion with the NS class.)}
    \label{fig:ns_prob_phot_ast}
\end{figure}

In this section, we discuss prospects for distinguishing NS lenses from other classes using their $\log_{10} t_{\rm E}$--$\log_{10}\theta_{\rm E}$ distributions. For each of the 4 simulation runs, we build a classifier returning the lens class probability for a given point or distribution in a subspace of observables as described in \citet{Kaczmarek2025}; we refer to that work for a detailed description of the classification method. In summary, we use the Bayes theorem to define the probability of the lens causing a microlensing event with a set of observable parameters $\phi$ belonging to $\text{class}_L$ under a Galactic model $\mathcal{G}$:

\begin{equation}
p(\text{class}_L| \phi, \mathcal{G}) = \frac{p(\text{class}_L| \mathcal{G})p(\phi| \text{class}_L, \mathcal{G})}{p(\phi| \mathcal{G})}.
\label{eq:bayes}
\end{equation}

We take $p(\text{class}_L| \mathcal{G})$ directly from the event counts per class in the simulated dataset. To construct continuous $p(\phi| \text{class}_L, \mathcal{G})$ from discrete sampling, we use Gaussian KDEs constructed on the $\phi$ parameter space with detectable simulated events of a given $\text{class}_L$ as implemented in {\tt scipy} \citep{scipy} with default settings. We note that Eq.~\eqref{eq:bayes} assumes perfect measurement of the observable parameters $\phi$, which makes this analysis optimistic. 
However, by the very nature of the spur-feature being in the high $\theta_E$ portion of the parameter space, the lenses of interest are expected to preferentially skew towards high signal to noise ratios, making this assumption less impactful than it would normally be for analyses focused on the bulk of the observed catalog. $p(\phi| \mathcal{G})$ is simply a normalising factor independent of $\text{class}_L$, which we apply by ensuring $\sum_{\text{class}_L} p(\text{class}_L| \phi,  \mathcal{G}) = 1$. \highlight{We make one change compared to \citet{Kaczmarek2025} in that we incorporate NS and BH weights in the creation of the respective KDEs.}

To account for the epistemic uncertainty of the classifier, in addition to the astrophysical classes \{Star, WD, NS, BH\}, we include the None class which is assigned when the parameters $\phi$ are not consistent with those of event populations in the underlying simulation (see Section 3.3 and Appendix A in \citealt{Kaczmarek2025} for a detailed explanation of the None class). The None class is constructed following the formula:

\begin{equation}
    p(\phi | \text{None},\mathcal{G}) = A \left( 1 - \frac{p(\mathbf{\phi} | \mathcal{G})}{\max_{\phi} (p(\mathbf{\phi} | \mathcal{G}))} \right)
    \label{eq:none},
\end{equation} 

\highlight{where $A$ is a constant setting the None class weight and $p(\mathbf{\phi} | \mathcal{G})$ is approximated with a tophat KDE constructed on the $\phi$ parameter space with all detectable simulated events, using the {\tt scikit-learn} \citep{Pedregosa2011} implementation. Here, we set $A$ to $10^{-4}$ and the KDE bandwidth to 0.3. As the None class is designed to represent parameters with no support in the simulation samples, the construction of this KDE is not impacted by NS/BH weighting.}

\begin{figure}
    \centering
    \includegraphics[width=\columnwidth, trim=0cm 0cm 0cm 0cm]{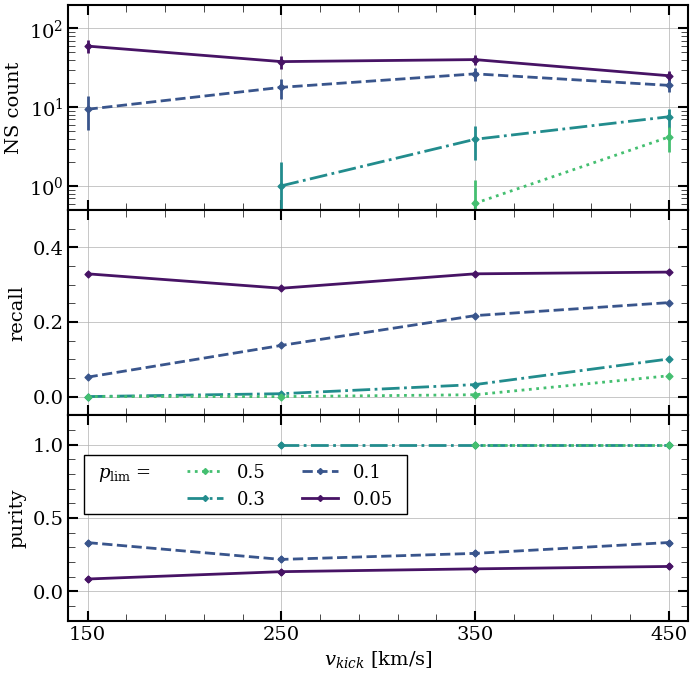}
    \caption{\textit{Top:} Count of retained NS events within an event sample constructed using a given $p_{\rm lim}$, rescaled to yields from the full GBTDS footprint. Errorbars representing Poisson noise as in Fig.~\ref{fig:ns_yields}. \textit{Center:} Recall of NS events, calculated as a ratio of selected NS events to all detectable NS events. \textit{Bottom:} Purity of the sample, calculated as a ratio of selected NS events to all events selected for the sample. For higher $p_{\rm lim}$, values for $v_{\rm kick} \in$ (150, 250) km/s are missing as samples cannot be constructed (i.e. the resulting sample size is 0).}
    \label{fig:purity_recall}
\end{figure}

In practice, we construct the KDEs on a 1000x1000 grid and evaluate $p(\text{class}_L| \phi, \mathcal{G})$ using the nearest gridpoint. Using this method, one can construct relative probability maps, i.e. plots showing what $p(\text{class}_L| \phi, \mathcal{G})$ would be assigned as a function of event parameters $\phi$ assuming they are known exactly. We show such probability maps for $v_{\rm kick} \in $ (150, 450) km/s in Fig.~\ref{fig:ns_prob_phot_ast}, demonstrating how moving from the space of photometry-only observables $(t_{\rm E}, \pi_{\rm E})$ to that of high-signal, photometry + astrometry observables $(t_{\rm E}, \theta_{\rm E})$ allows for high-confidence NS classification in case of high $v_{\rm kick}$. For the photometry-only subplots, we use all events passing only the magnitude cut and the photometrically detectable signal cut, whereas for the photometry+astrometry subplots, we use events passing all Subsection~\ref{sec:postproc} cuts including also the astrometrically detectable signal cut.

\highlight{To test prospects for constructing NS samples with this classifier, we use threshold probabilities $p_{\rm lim}$: we include a simulated event with parameters $\phi$ in our sample if  $p(\text{class}_L| \phi, \mathcal{G}) > p_{\rm lim}$. We construct such samples for all simulation runs and for several thresholds: $p_{\rm lim} \in (0.05, 0.1, 0.3, 0.5)$. We then correct our included and total event counts for NS and BH weights. We illustrate the count of retained NS events, recall and purity in this experiment in Fig.~\ref{fig:purity_recall}.}

\highlight{Samples constructed using a very low $p_{\rm lim} = 0.05$ have consistently high ($0.3-0.4$) recall, but at a cost of low ($0.1-0.2$) purity. Still, we note that this fraction of NS within the sample is 1-2 orders of magnitude larger than in the entire dataset. Those samples contain several 10s of NS and are selected from the `spur' and surrounding regions. Conversely, samples constructed using $p_{\rm lim} \in \{0.3, 0.5\}$ contain only NS, but at a cost of low recall ($<$10 events). Those samples are selected from the far side of the `spur'; for lower $v_{\rm kick}$, they cannot be constructed as there is no region of the parameter space sufficiently unique to NS.}

We note that belonging to the NS lens class can be verified independently -- e.g., with the lens mass being consistent with a NS mass and the non-detection of additional blended light expected from a luminous (e.g., main-sequence) star of the determined lens mass and distance. Therefore, the above exercise can be interpreted as constructing a sample of promising candidates that should be allocated additional analysis and observation resources, similarly to BH lens candidates in \citet{Kaczmarek2025}. In this situation, thresholds should be determined by adjusting the sample size to the resources available.

\section{Discussion and conclusions}
\label{sec:discussion}

We have conducted simulations of gravitational microlensing events tailored to the most up-to-date proposed \textit{Roman Space Telescope} Galactic Bulge Time Domain Survey design. We focused specifically on neutron star (NS) lenses, which are very hard to identify with the instrumentation constraints of current microlensing surveys due to confusion with other lens classes, but will become discoverable with \textit{Roman}. To assess the impact of NS natal kicks on microlensing observables, we ran four simulations with Maxwellian kick distributions, varying the mean kick velocity $v_{\rm kick}$.

We find that distributions of microlensing observables differ for NS populations with different $v_{\rm kick}$. In the $\log_{10} t_{\rm E}$--$\log_{10}\theta_{\rm E}$ space, which is favourable for remnant classification, we find a feature that is uniquely characteristic of NS lenses: a spur diagonally offset from the main distribution towards lower $t_{\rm E}$ and higher $\theta_{\rm E}$. The strength and exact location of this structure varies with $v_{\rm kick}$, making it most distinct at the high end, $v_{\rm kick} = 450$ km/s.

\highlight{We have identified the `spur' NS as \textit{mesolenses} --  a class of nearby, fast objects that have a high probability of causing lensing events and may even cause repeated events. Large kicks cause both effects strengthening the spur (increased potential for generating mesolenses) and weakening it (decreased NS density), making it especially challenging to study this structure and use it for drawing constraints on kick distributions. Detailed studies including careful modelling of NS dynamics, higher sample sizes, and more kick velocity distributions are needed.}

\review{We note that if the NS spur feature is detectable in real Roman observations, it could enable the identification of highly likely NS candidates without requiring a $\pi_{E}$, which may represent a
bottleneck for full lens characterization in GBTDS microlensing events. We therefore recommend that all events found in this region should be treated with special care and allocated necessary follow-up resources by prioritizing their ingestion into Target and Observation management systems \citep[e.g.,][]{Street2018, Coulter2023} so they can be tracked and this new potential characterization pathway can be investigated.
}

We predict \romanst~will find approximately 11000 microlensing events detectable in both photometry and astrometry, \highlight{$\sim100$ of which will be} isolated neutron stars. With direct mass measurements available via a complete resolution of microlensing parameters, this dataset will provide outstanding statistical constraints on the NS equation of state and the speculated mass gap. In total, this dataset will contain approximately \highlight{2500} compact objects, providing an unprecedented census of isolated stellar remnants. While in this work we focus on NS classification and population properties, we note the GBTDS-tailored simulation results will also be useful for other science cases, and make them publicly available within the {\tt popclass} classification software.

We note this study is constrained by several limitations. All predictions about \romanst~-- concerning both survey strategy and instrument performance -- are preliminary and will need to be verified in the later stages of the mission. Furthermore, a simulation can only be as accurate as the underlying Galactic model. {\tt PopSyCLE} relies on state-of-the-art models of the Milky Way and stellar evolution. However, some questions about the Galactic structure remain open, including bar angle and pattern speed, mass distribution, etc. \citep[e.g.,][]{HuntVasiliev2025}. There are even more unknowns regarding the creation of remnants. We assume a single Maxwellian distribution for natal kicks, while more complex distributions are also discussed -- e.g. \citet{Igoshev2020} propose a bimodal kick distribution. The initial-final mass relation (IFMR) between progenitors and remnants is not well-known; it is difficult to test (especially in the mass-gap range) due to the obstacles and biases in remnant detection and mass measurement discussed in Section~\ref{sec:intro}. Various initial-final mass prescriptions, using different methods, have been proposed in the literature (see \citealt{Rose2022} for a detailed description of IFMRs integrated into {\tt PopSyCLE} and their impact on microlensing).

\highlight{Notably, \citet{Sweeney2022} have demonstrated with simulations that the spatial Galactic distribution of NS should be more diffuse than that of other astrophysical classes, due to both their progenitors experiencing the evolving structure of the Galaxy and their changed post-kick orbits. However, {\tt PopSyCLE} currently does not model the latter. For more discussion of the impact of this limitation on microlensing predictions, we refer to \citet{Lam2020}.}

\highlight{We attempt to correct for this effect in post-processing. Namely, we run dynamical simulations for $10^5$ simulated pre-kick NS sampled from the \citet{Sweeney2022} dataset to obtain four post-kick, present-day Galactic distributions of NS.  We use the {\tt StellarMortis} software \review{\citep{software:StellarMortis}} developed as part of the \citet{Sweeney2022} work. We approximate post-kick to pre-kick density ratios for a given $v_{\rm kick}$ and Galactocentric position by bin count ratios. We then assign weights to our simulated NS correcting for the changed densities. This correction lowers the total yields by a factor of 1.4 (lowest $v_{\rm kick}$) to 5.3 (highest $v_{\rm kick}$).}

\highlight{Our treatment of massive remnants is similar to that of most recent microlensing studies dealing with black hole natal kicks. \citet{Koshimoto2024} account for the changed remnant distributions by modifying the Galactic model of \citet{Koshimoto2021} with analytical scale height and surface density profiles fitted to the numerical results of \citet{Tsuna2018}. They then generate black hole lenses following the prescription (IFMR and adding kicks) of \citet{Lam2020}. They find a similar trend of increasing black hole yields (there represented as a fractional contribution per $t_{\rm E}$ bin) with increasing $v_{\rm kick}$ when not applying disk inflation, but decreasing after this correction to the model. At the final stages of preparation of this work, a pre-print by \citet{Wu2025} has been made available which discusses black hole selection and event rates from a combination of ground-based microlensing surveys and interferometry. \citet{Wu2025} correct for black hole kick effects by applying weight factors to their expected event rates, separately for the disk (where they follow the analytical formulae of \citealt{Koshimoto2024}) and the bulge (which they model as an exponential distribution whose effective radius changes with kick velocity, assuming a steady state).}

\highlight{We note that the solution of applying weights used in this work is not perfect. While the first-order effect of dynamical evolution is the changed NS density, we also expect velocity distributions to be impacted as NS orbits change. This effect is very difficult to model and cannot be handled with a simple correcting factor, as present-day NS velocity distributions are expected to vary with Galactic position and age (see e.g., Fig. 3 in \citealt{Sweeney2022}). Clearly, it would be ideal to have a fully realistic population of NS with present-day positions and velocities available to use as lenses. However, to create such a population, one would need to consider that the NS are created in a different Galactic position -- i.e., at a different line of sight -- than they are observed as lenses. This is incompatible with the way {\tt PopSyCLE} works, which includes computationally expensive population synthesis of stars along a chosen line of sight to simulate small circular fields (see e.g. Fig.~\ref{fig:footprint}). A full simulation including the dynamical evolution of NS across all possible lines of sight, i.e. the entire Galaxy, would be extremely computationally expensive. For example, even without simulating microlensing, \citet{Sweeney2022} needed to apply a downsampling of remnant and star populations by $10^3$ and $10^6$, respectively, to make the aforementioned study computationally feasible. Such a downsampling would be far from sufficient for studying parameter distributions or estimating yields for a survey like GBTDS. {\tt PopSyCLE}, considering all NS in a given pencil beam with no downsampling, only returns $\sim10^2$ NS lens yields in our simulations. Finally, we note that our weighting factors assume axisymmetry; indeed, our simulated samples of post-kick NS are well-described by an axisymmetric distribution, as  \citet{Sweeney2022} and {\tt StellarMortis} do not include non-axisymmetric components (e.g. bar) of the Milky Way potential.}

\highlight{However, as high-performance clusters and simulation codes develop exponentially, it is conceivable that microlensing simulations may no longer be hindered by these computational limitations in the near future. We have outlined some areas of focus that might be useful for such a future study. Other limitations discussed above may also be alleviated in the coming years with the start of \romanst~operations, Galactic models informed by new surveys including the upcoming \textit{Gaia} data releases, and improvements in stellar evolution codes.}

Regardless of these limitations, we make promising observations on the \romanst~isolated NS population\highlight{. }We find a feature in the observable parameter space that is characteristic to NS. The shape and strength of this feature is dependent on NS kicks. Hence, it can be leveraged to constrain natal kick distributions. \highlight{Similarly, the distance distribution of NS lenses is also dependent on NS kicks. We find that large kicks exhibit effects that both strengthen (via high NS velocities) and weaken (via decreased NS density, especially at low $|z|$) these characteristic features. The interplay of these effects should be treated with special care and ideally studied with dedicated simulations focusing on NS dynamics in the future.}


\highlight{\review{In the course of this work, we produced simulations extending beyond the neutron star population}. We \review{also extended} the {\tt popclass} classification software \citep{popclass} to include our simulation results in anticipation of GBTDS. We provide expected yields of events detectable by \romanst~for each of the astrophysical lens classes (stars, white dwarfs, neutron stars and black holes). We demonstrate the impact of survey strategy and detectability criteria on event yields, and we make all simulated events available with detectability flags so that these criteria can be further analysed and adjusted. All microlensing event datasets generated in this study \highlight{will be made publicly available upon the acceptance of this paper}.} 

\begin{acknowledgements}
The authors acknowledge support by the state of Baden-Württemberg through bwHPC. We would like to thank Joachim Wambsganss, David Sweeney, Oskar Grocholski, Jessica Lu, Macy Huston, and Natasha Abrams for useful discussions. \review{We thank the anonymous reviewer for their comments that helped improve the quality of this paper.} This work was performed under the auspices of the U.S. Department of Energy by Lawrence Livermore National Laboratory under Contract DE-AC52-07NA27344. The document number is LLNL-JRNL-2013705. This work was supported by the LLNL LDRD Program under Project 22-ERD-037.
\end{acknowledgements}

%
%

\bibliographystyle{aa}
\bibliography{refs}

\end{document}